# Numerical Study of Mechanism of U-shaped Vortex Formation

Ping Lu[1], Qin Li[2] and Chaoqun Liu[3]

UNIVERSITY OF TEXAS AT ARLINGTON, ALRINGTON, TX 76019, USA
CLIU@UTA.EDU

This paper illustrates the mechanism of U-shaped vortex formation which is found both by experiment and DNS. The main goal of this paper is to explain how the U-shaped vortex is formed and further develops. According to the results obtained by our direct numerical simulation with high order accuracy, the U-shaped vortex is part of the coherent vortex structure and is actually the tertiary streamwise vortices induced by the secondary vortices. The new finding is quite different from existing theories which describe that the U-shaped vortex is newly formed as the head of young turbulence spot and finally break down to small pieces. In addition, we find that the U-shaped vortex has the same vorticity sign as the original λ-shaped vortex tube legs and serves as a second neck to supply vorticity to the ring-like vortex when the original vortex tube is stretched and multiple rings are generated.

## Nomenclature

$M_\infty$ = Mach number
$Re$ = Reynolds number
$\delta_{in}$ = inflow displacement thickness
$T_w$ = wall temperature
$T_\infty$ = free stream temperature
$Lz_{in}$ = height at inflow boundary
$Lz_{out}$ = height at outflow boundary
$Lx$ = length of computational domain along x direction
$Ly$ = length of computational domain along y direction
$x_{in}$ = distance between leading edge of flat plate and upstream boundary of computational domain
$A_{2d}$ = amplitude of 2D inlet disturbance
$A_{3d}$ = amplitude of 3D inlet disturbance
$\omega$ = frequency of inlet disturbance
$\alpha_{2d}, \alpha_{3d}$ = two and three dimensional streamwise wave number of inlet disturbance
$\beta$ = spanwise wave number of inlet disturbance
$R$ = ideal gas constant
$\gamma$ = ratio of specific heats
$\mu_\infty$ = viscosity

## I. Introduction

THE transition process from laminar to turbulent flow in boundary layer is a basic scientific problem in modern fluid mechanics and has been the subject of study for over a century. Many different concepts for the explanation of the mechanisms involved have been developed based on numerous experimental, theoretical, and numerical investigations. After over a century study on the turbulence, the linear and early weakly non-linear stages of flow transition are pretty well understood. However, for late non-linear transition stages, there are still many questions remaining for research (Kleiser et al, 1991; Borodulin et al, 2002; Bake et al 2002, Rist et al, 2002; Kachanov, 2003). U-shaped vortex was found both by experiment (Borodulin et al, 2002) as "barrel shaped wave" and by DNS (Singer et al, 1994) as newly formed vortex which plays a role as the head of young turbulence spot and eventually broke down to small pieces. In order to get deep understanding on the late flow transition in a boundary layer, we recently conducted a high order DNS with 1920×241×128 gird points and about 600,000 time steps to study the mechanism of the late stage of flow transition in a boundary layer. Many new findings are made and new mechanisms are revealed (Chen et al 2010a, 2010b, 2010c; Liu et al, 2010a, 2010b, 2010c, 2010d). We have a number of new findings on late flow transition in a boundary layer: 1) the widely spread concept "hairpin vortex breakdown to small pieces", which was considered as the last step of flow transition, is not found in this study (Figure 8); 2) the ring-like vortex is found the only form existing inside the flow field; 3) the ring-like vortex formed from the $\Lambda$-vortex is the result of the interaction between two pairs of counter-rotating primary and secondary streamwise vortices; 4) the formation of the multiple ring structure follows the first Helmholtz vortex conservation law. The primary vortex tube rolls up and is stretched due to the velocity gradient. In order to maintain vorticity conservation, a bridge must be formed to link two $\Lambda$-vortex legs. The bridge finally develops as a new ring. This process keeps going on to form a multiple ring structure (Figure 8); 5) the U-shaped vortices are not newly generated secondary vortices, but are existing as a coherent vortex structure; 6) neither the hairpin vortex nor U-shaped vortex breaks down (Figure 21). The previous reports on vortex breakdown were either based on 2-D visualization or using low pressure center as the vortex tube (Figure 7); 7) the small vortices can be found on the bottom of the boundary layer near the wall surface(bottom of boundary layer). It is believed that the small vortices, and thus turbulence, are all generated by high shear layer and the wall surface instead of by "large vortex breakdown".

---

[1] PhD Student & AIAA member, University of Texas at Arlington, USA
[2] Post Doctor, University of Texas at Arlington, USA
[3] Professor & AIAA Associate Fellow, University of Texas at Arlington, USA

In this paper, a new study for the mechanism of the U-shaped vortex formation is presented. The U-shaped vortex is found as part of the coherent vortex structure. It is induced by the secondary vortices, which has the same vorticity sign as the original $\lambda$-shaped vortex tube legs, and serves as a second neck to supply vorticity to the multiple rings when the original vortex tube is stretched and multiple rings are generated. Here, we use the $\lambda_2$ criterion (Jeong & Hussain, 1995) for visualization.

## II. Numerical Simulations

**2.1 Case setup**

The computational domain is displayed in Figure 1. The grid applied is 1920×128×241, representing the number of grids in streamwise ($x$), spanwise ($y$), and wall normal ($z$) directions. The grid is stretched in the normal direction and uniform in the streamwise and spanwise directions. The length of the first grid interval in the normal direction at the entrance is found to be 0.43 in wall units ($Y^+$=0.43).

The parallel computation is accomplished through the Message Passing Interface (MPI) together with domain decomposition in the streamwise direction. The computational domain is partitioned into N equally-sized sub-domains along the streamwise direction. N is the number of processors used in the parallel computation. The flow parameters, including Mach number, Reynolds number etc are listed in Table 1. Here, $x_{in}$ represents the distance between leading edge and inlet, $Lx$, $Ly$, $Lz_{in}$ are the lengths of the computational domain in x-, y-, and z-directions, respectively, and $T_w$ is the wall temperature. $\delta_{in}$ is the momentum thickness of the boundary at the inlet.

| $M_\infty$ | $Re$ | $x_{in}$ | $Lx$ | $Ly$ | $Lz_{in}$ | $T_w$ | $T_\infty$ |
|---|---|---|---|---|---|---|---|
| 0.5 | 1000 | 300.79$\delta_{in}$ | 798.03$\delta_{in}$ | 22$\delta_{in}$ | 40$\delta_{in}$ | 273.15K | 273.15K |

**Table 1. Flow parameters**

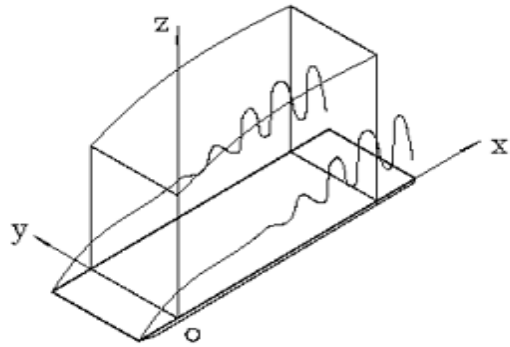

**Figure 1. Computation domain**

**2.2 Code validation and DNS results visualization**

The skin friction coefficient calculated from the time-and spanwise-averaged profile is displayed in Figure 2. The spatial evolution of skin friction coefficients of laminar flow is also plotted out for comparison. It is observed from these figures that the sharp growth of the skin-friction coefficient occurs after $x \approx 450\delta_{in}$, which is defined as the 'onset point'. The skin friction coefficient after transition is in good agreement with the flat-plate theory of turbulent boundary layer by Cousteix in 1989 (Ducros, 1996).

Time- and spanwise- averaged streamwise velocity profiles for various streamwise locations, which are obtained by a coarse grid (960×64×121) and a fine grid (1920×128×241), are shown in Figure 3. The inflow velocity profiles at $x=300.79\delta_{in}$ is a typical laminar flow velocity profile. At $x = 632.33\delta_{in}$ the mean velocity profile approaches a turbulent flow velocity profile.

The code validation clearly shows that we obtained the turbulent flow profile and our calculation is correct. The agreement between the coarse and fine grids also shows we obtained a grid convergence.

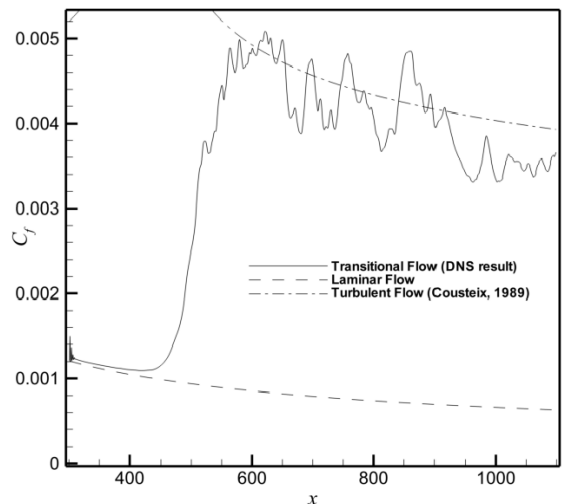

**Figure 2. Streamwise evolution of the time-and spanwise-averaged skin-friction coefficient**

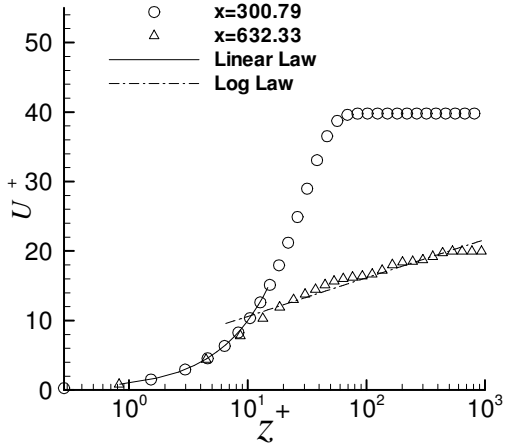 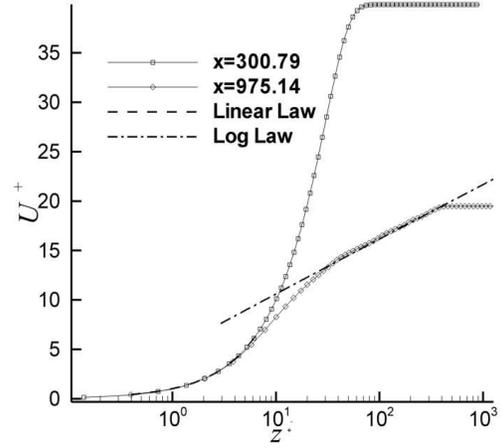

**(a) Coarse grids (960×64×121)**        **(b) Fine grids (1920×128×241)**

**Figure 3. Log-linear plots of the time-and spanwise-averaged velocity profile in wall unit**

In boundary layer flows, viscosity is non-negligible. Standard approaches, such as integrating vortex lines, using minimum pressure or maximum vorticity, may lead to improper vortex identification. Jeong and Hussain have established a robust criterion for identification of vortex (or coherent) structures in viscous flows based on the eigenvalues of the symmetric $3\times 3$ tensor

$$M_{ij} := \sum_{k=1}^{3} \Omega_{ik}\Omega_{kj} + S_{ik}S_{kj} ,  \tag{1}$$

where

$$\Omega_{ij} := \frac{1}{2}(\frac{\partial u_i}{\partial x_j}+\frac{\partial u_j}{\partial x_i}) \text{ and } S_{ij} := \frac{1}{2}(\frac{\partial u_i}{\partial x_j}-\frac{\partial u_j}{\partial x_i}) \tag{2}$$

represent the symmetric and anti-symmetric components of the velocity gradient tensor, $\nabla u$. Given the three (real) eigenvalues of M at each grid point, a vortex core is identified as any contiguous region having two negative eigenvalues. If the eigenvalues are sorted such that $\lambda_1 \leq \lambda_2 \leq \lambda_3$, then any region for which $\lambda_2 < 0$ corresponds to a vortex core. One advantage of this approach is that vortices can be identified as isosurfaces of a well-defined scalar field. Moreover, the criterion $\lambda_2(x) < 0$ is scale invariant, so in principle there is no ambiguity in selecting which isosurface value to render. In practice, one usually biases the isosurface to a value that is below zero by a small fraction of the full dynamic range in order to avoid noise in regions where the velocity is close to zero.

**2.3 Second Sweep and positive spike**

From the papers we have published, the second sweep starts from the center of the ring head and brings high speed flow down towards the wall (Figure 4). The second sweep, combining with the first sweep which is generated by the original $\lambda$ vortex legs, will form a strong positive spike (red color) in contrast to the negative spike (blue color) (see Figure 5). Figure 6 shows a strong secondary vortex which is formed near the wall.

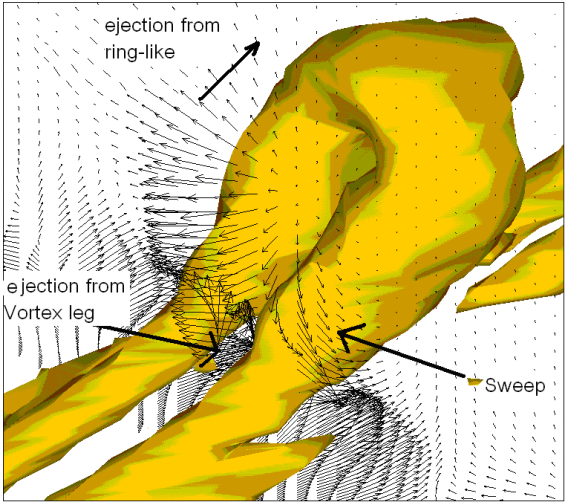 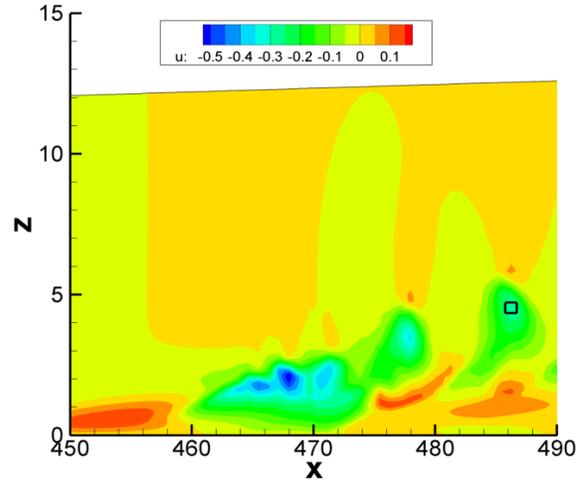

**Figure 4. Second sweep**  **Figure 5. Negative and positive spike**

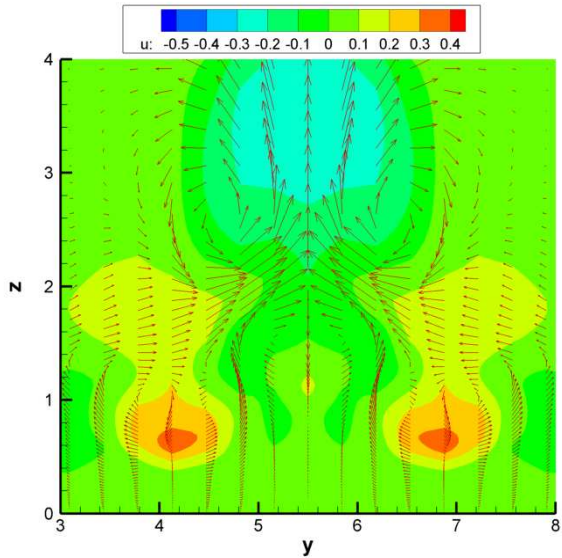

**Figure 6. Positive spike**

### 2.4 U-shaped vortex formation and roles

U-shaped vortex was reported by Singer et al (1994) (see Figure 7) who believed that the U-shaped vortex was newly generated and broke down to small pieces. However, according to the results obtained by our new DNS, the U-shaped vortex actually exists for quite a long period and is part of the coherent vortex structure (Figure 8) around all the ring-like vortices except for the second ring-like vortex. According to our DNS observation, the U-shaped vortex never breaks down. The problem of the visualization presented by Singer and other literatures about vortex breakdown was caused by using 2-D visualization or using the low pressure center as the vortex center, which is not always true. The instantaneous pictures are drawn using $\lambda_2$ at time step t=8.0T with small negative value selected for visualization.

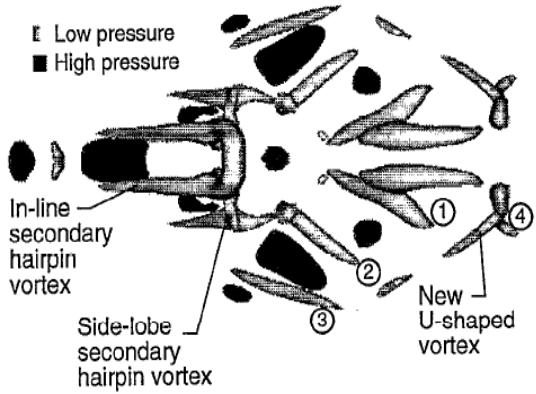

**Figure 7. U-shaped vortex (Singer, 1994)**

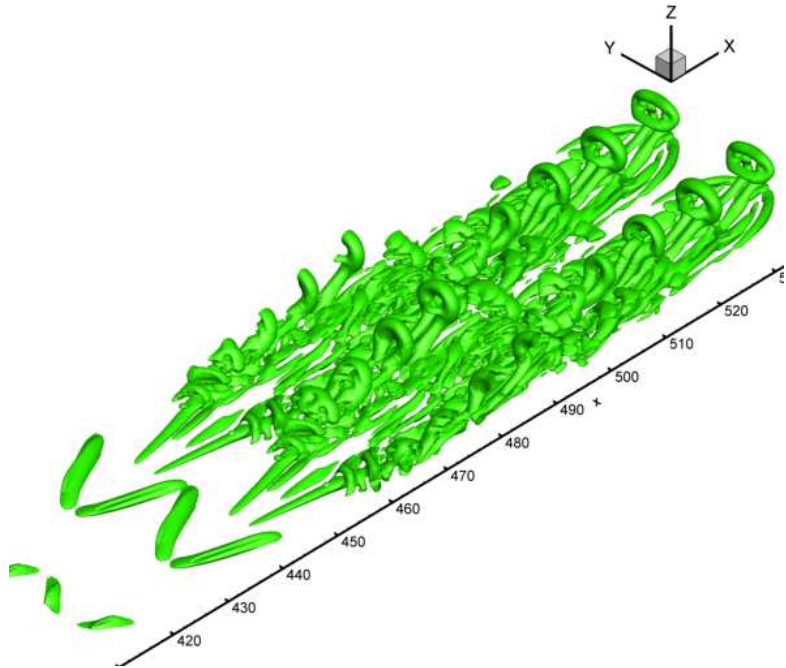

**Figure 8. U-shaped vortex is part of coherent vortex structure**

As shown in figure 9, the prime streamwise vortices are stretched and raised up, and then the streamwise voticity is formed by the interaction of the prime vortex and solid wall. The vorticity on the wall will further develop into secondary streamwise vortices by flow separation. From the $\lambda_2$ contour map and vectors at the section of x= 530.348 $\delta_{in}$ shown in Figure 10, we can find that the prime streamwise vorticity will create anti-rotational secondary streamwise vorticity because of the effect of the solid wall. The secondary streamwise vorticity is strengthened and the vortex detaches from the solid wall gradually. When the secondary vortex detaches from the wall, it will induce a new streamwise vorticity by the interaction of the secondary vortex and the solid wall which actually is a tertiary streamwise vortex. The tertiary vortex is the U-shaped vortex.

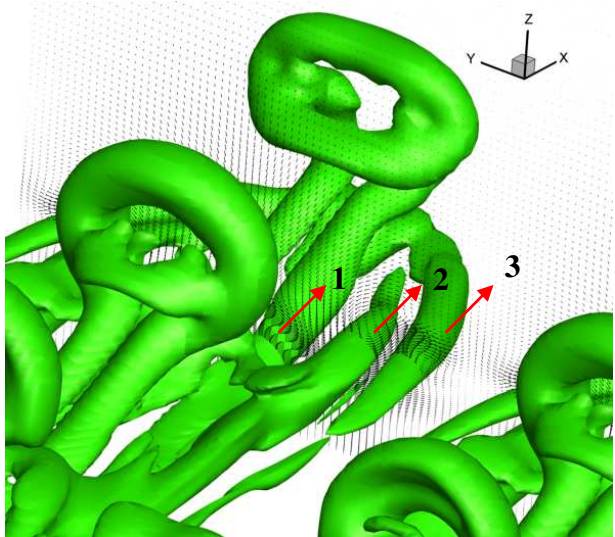 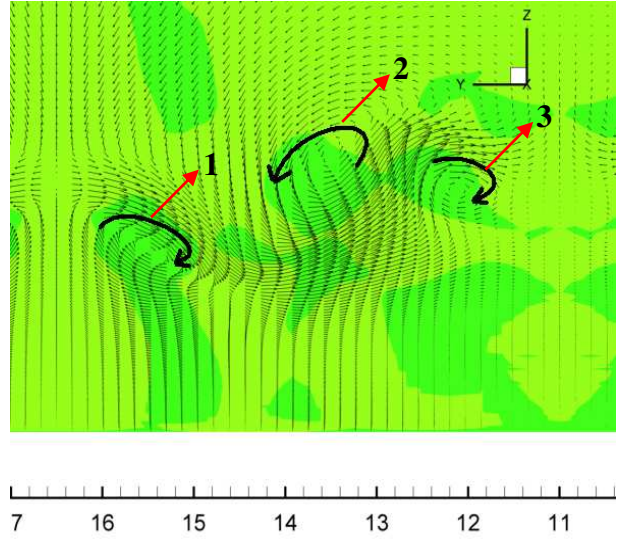

**Figure 9. isosurface of $\lambda_2$ and streamtrace at x= 530.348 $\delta_{in}$ (1 and 3-clockwise, 2-counterclockwise)**

**Figure 10. $\lambda_2$ contour and vector at section x=530.348 $\delta_{in}$**

Meanwhile, three sections (x=529.437 $\delta_{in}$, 530.348 $\delta_{in}$, 531.237 $\delta_{in}$) are selected to show the structures of vortices. From the Figures 12-15, we can see that the $\Lambda$-vortex is close to the wall and the secondary vortices are a little far from the wall. The U-shaped vortices are located near the wall in normal direction but their two legs are located far away from each other to form a U-shaped spot bound. From the analysis of the stream trace, the vorticity of the U-shaped vortex seems to have same signs as the leg of original $\Lambda$-vortex legs (clockwise). The first U-shaped structures can be clearly visualized at the x coordinates starting at 525 $\delta_{in}$ and ending at 537 $\delta_{in}$, the third one starting at 511 $\delta_{in}$ and ending at 518 $\delta_{in}$.

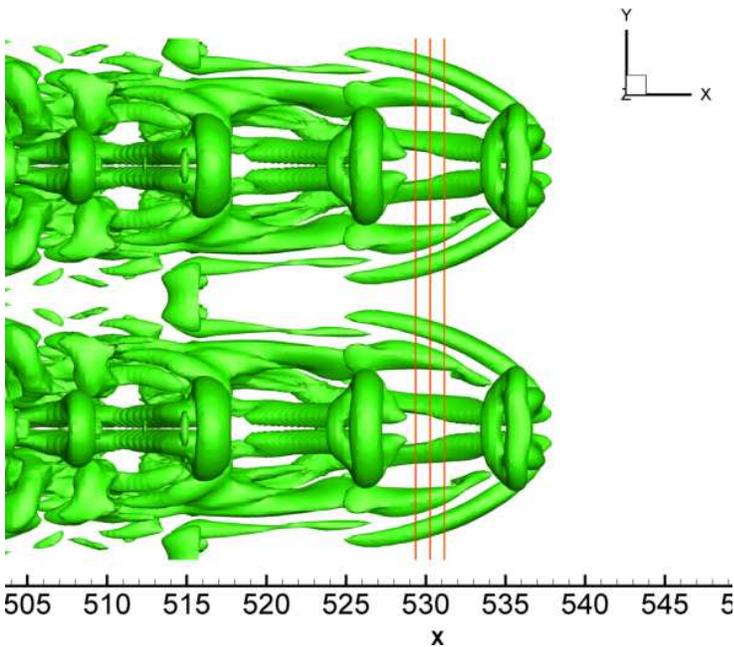

**Figure 11. Three slices along the streamwise vortex**

In order to reveal the relationship between original λ vortex legs, secondary vortex and U-shaped vortex, the streamtrace is drawn for each section to show the direction of each vortex. Figures 12 to 15 show that the sign of the U-shaped vortex is clockwise-rotating, which is the same as one of the original λ vortex legs. However, the sign of U-shaped vortex is different from the secondary vortex which is counterclockwise-rotating.

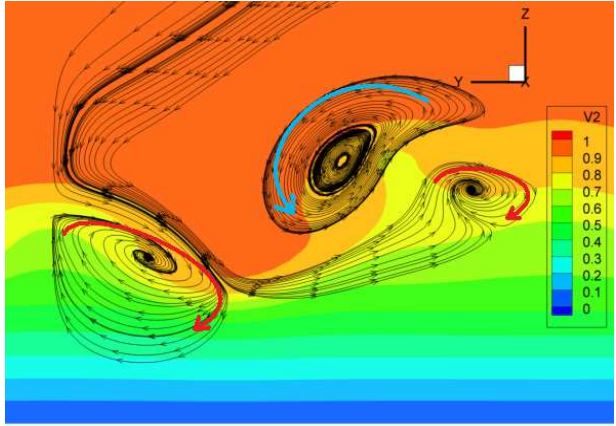
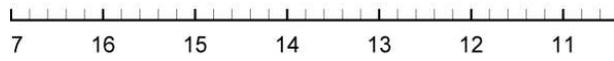

Figure 12. velocity contour and streamtrace at section x=529.437 $\delta_{in}$

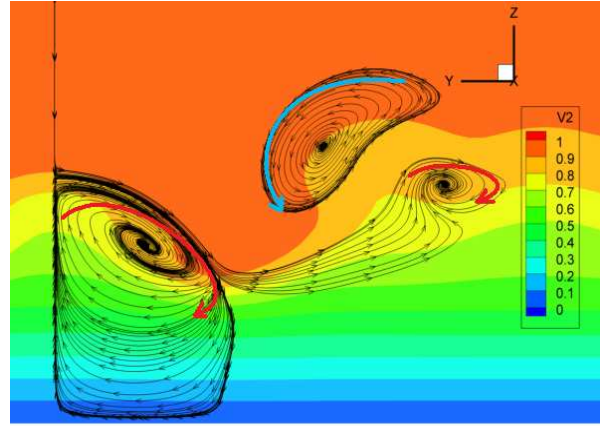
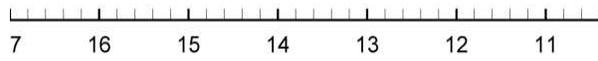

Figure 13. velocity contour and streamtrace at section x=530.348 $\delta_{in}$

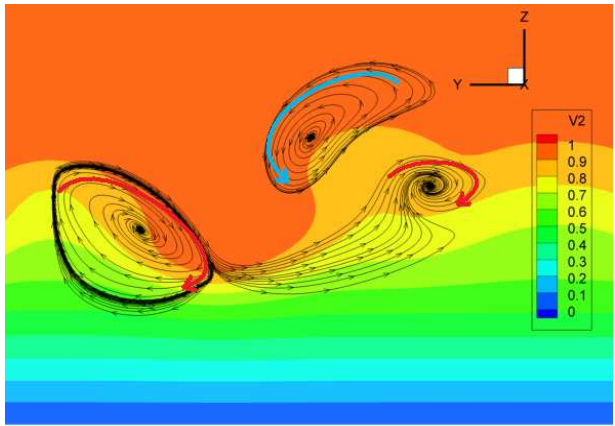
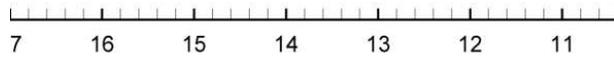

Figure14. velocity contour and streamtrace at section x=531.237 $\delta_{in}$

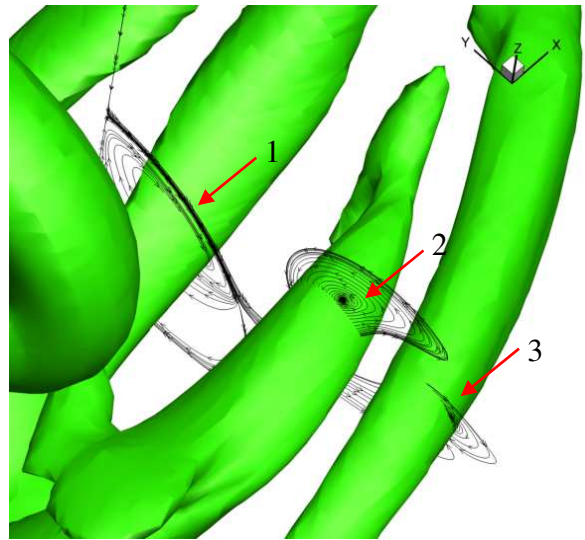

Figure 15. isosurface of $\lambda_2$ and streamtrace at section x=530.348 $\delta_{in}$
(1 and 3-clockwise, 2-counterclockwise)

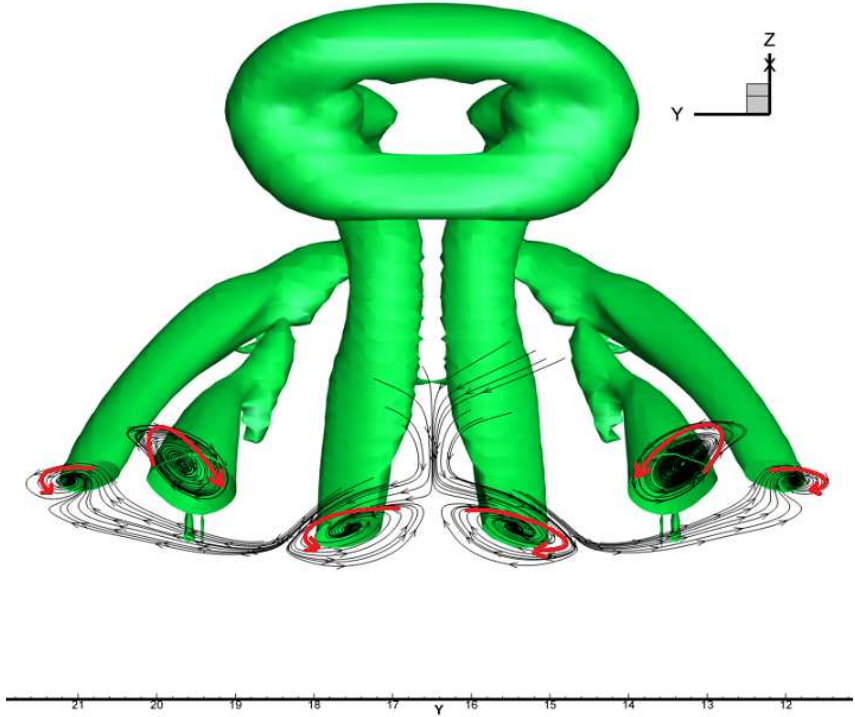

**Figure 16. isosurface of $\lambda_2$ and streamtrace U-shaped vortex**

Since the U-shaped vortices play a vital role in the formation of multiple rings, it is very important to study their mechanism. Analysis of the vorticity $\Omega_x$ shows that this U-shaped structure consists mostly of streamwise vorticity, and it is aligned primarily in the streamwise direction. From the observation of our computation running at very late stage, the $\lambda$ ('horseshoe') vortices travel for a long distance, becoming weaker and weaker, but they never break down. One of the reasons is that the U-shaped vortex serves as a second neck to supply vorticity to the multiple rings in order to keep the vorticity conservation. Between the ring-like vortices and the U-shaped vortices, there is a strong connection that when both $\lambda$ and U-shaped vortex tubes are close to each other, they will merge together (Figures 17, 18, 19 and 20). Note that they have the same sign of vorticity.

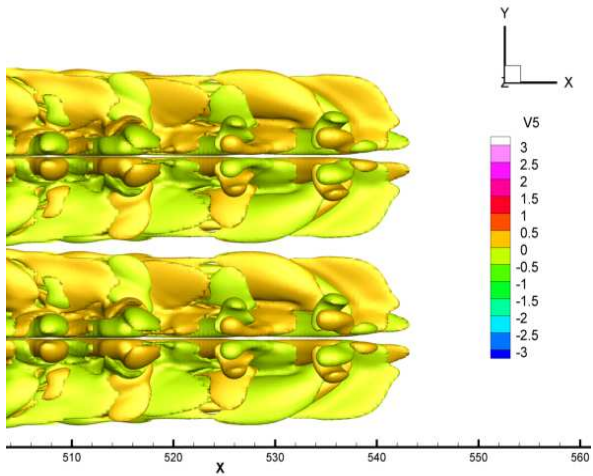 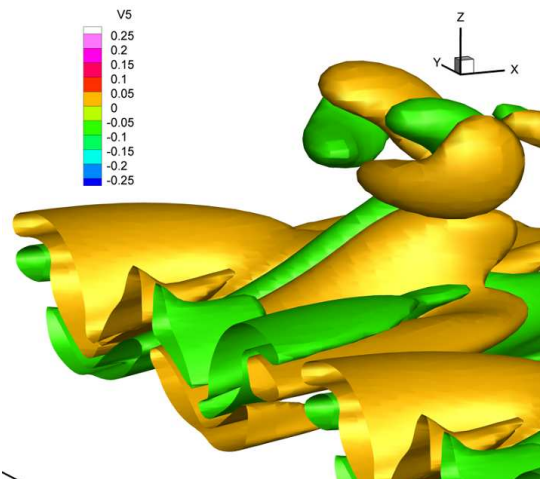

**Figure 17. contour of vorticity $\Omega_x$**  **Figure 18. vorticity $\Omega_x$ of first U-shaped vortex**

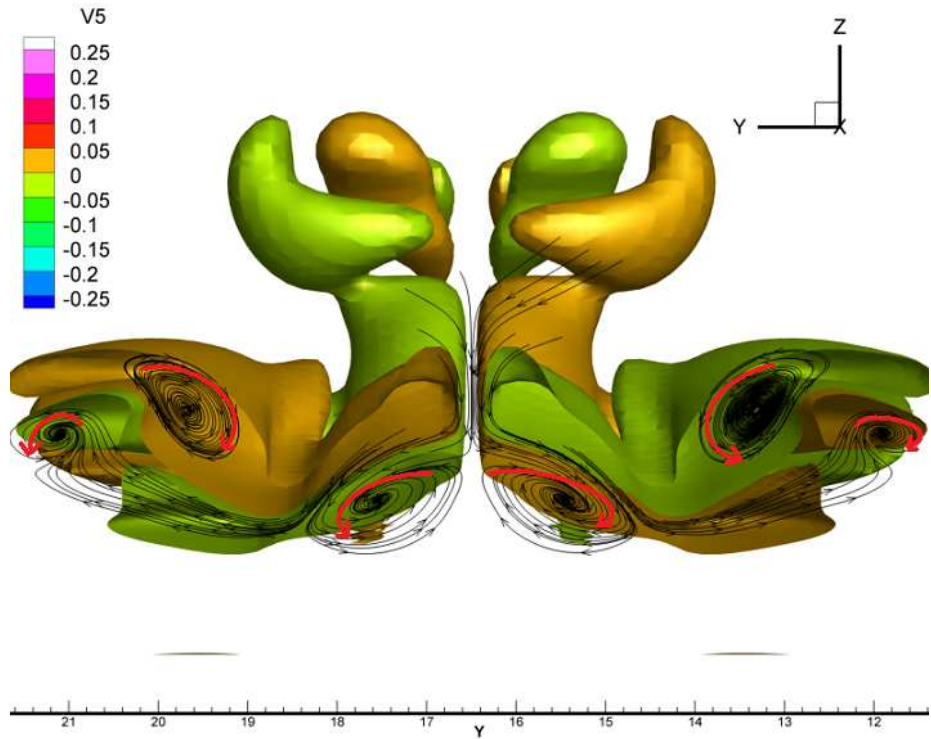

**Figure 19. vorticity $\Omega_x$ distribution and streamtrace slice cutting at x=529.191**

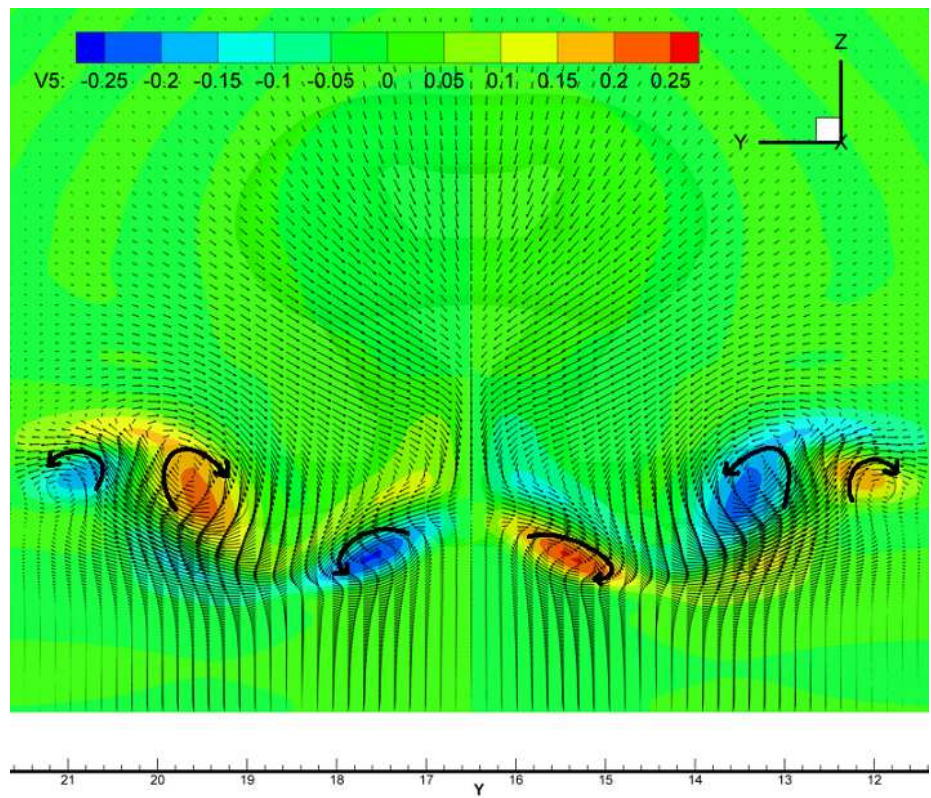

**Figure 20. contour of vorticity $\Omega_x$ and streamtrace**

Based on our numerical simulation, the U-shaped vortex exists from the beginning of the multiple vortex structure formation and lies underneath the original $\lambda$-vortex legs. However, the U-shaped vortex only becomes clear around the heading ring at the late transition stage. For other rings, the U-shaped vortex is surrounded by many small length scales and is hard to be distinguished. The reason is that when the ring reaches the inviscid zone, it cannot rise up anymore. It also cannot break away from the vortex leg. In that case, the vertical position is the most stable position. As the head of the leading ring-like vortex skews (not perfectly circular) and slopes (not perpendicular), the second sweep and thus positive spike die. The small length scales disappear quickly and are damped completely by viscosity in the boundary layer. Since the U-shaped vortex is induced by the secondary vortex and has no relation with the positive spike, it becomes clear as the small length scales are damped. The U-shaped vortex is located beside the original $\lambda$-vortex legs. The distance of two legs of the U-shaped vortex is much larger than the original $\lambda$-vortex legs (Figure 21).

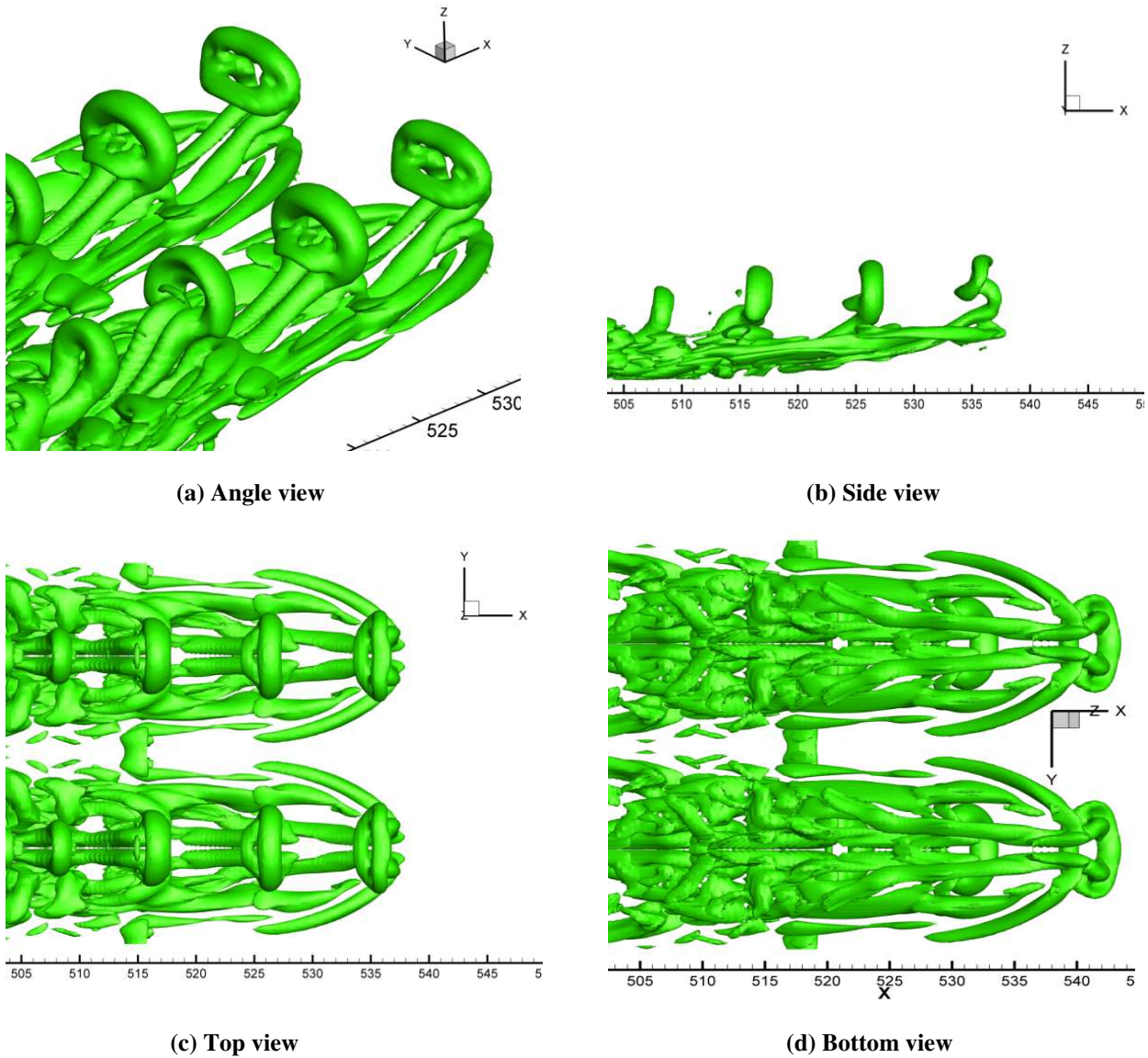

(a) Angle view  (b) Side view

(c) Top view  (d) Bottom view

**Figure 21. View of young turbulence spot head from different directions (t=8.0T)**

## III. Conclusions

Based on our new DNS study, the following conclusions can be made:

1. U-shaped vortex is part of the coherent vortex structure and exists from the beginning multiple ring generation. It is not newly generated.
2. U-shaped vortex is a tertiary vortex induced by the secondary vortex and has much larger distance between two legs than the distance between the original $\lambda$-vortex legs.
3. U-shaped vortex has the same sign as the original $\lambda$-shaped vortex legs.
4. As the multiple rings are generated, the vortex ring heads become weaker and weaker according to the vorticity conservation. However, the U-shaped vortex serves as a second neck to supply vorticity to the rings. This is one of the reasons why the multiple ring structure can travel for a long distance.
5. U-shaped vortex becomes clear (Figure 21) because the heading ring is skewed and sloped. The second sweep becomes weak and positive spike disappears. The small length scales are damped quickly by the boundary layer viscosity due to the lack of energy supplies by second sweep from the invicid region.

## Acknowledgments

This work was supported by AFOSR grant FA9550-08-1-0201 supervised by Dr. John Schmisseur. The authors are grateful to Texas Advanced Computing Center (TACC) for providing computation hours.